\documentclass[twocolumn,prb,showpacs,preprintnumbers,amsmath,amssymb,floats,superscriptaddress]{revtex4-1}
\usepackage[latin9]{inputenc}
\usepackage{color}
\usepackage{amsmath}
\usepackage{amssymb}
\usepackage{stmaryrd}
\usepackage{graphicx}
\usepackage{esint}
\usepackage{lineno}
\usepackage{fixmath}

\makeatletter

\@ifundefined{textcolor}{}
{%
 \definecolor{BLACK}{gray}{0}
 \definecolor{WHITE}{gray}{1}
 \definecolor{RED}{rgb}{1,0,0}
 \definecolor{GREEN}{rgb}{0,1,0}
 \definecolor{BLUE}{rgb}{0,0,1}
 \definecolor{CYAN}{cmyk}{1,0,0,0}
 \definecolor{MAGENTA}{cmyk}{0,1,0,0}
 \definecolor{YELLOW}{cmyk}{0,0,1,0}
}

\usepackage{epstopdf}
\usepackage{color}
\usepackage{lipsum}
\usepackage{dcolumn}
\usepackage{bm}

\makeatother

\begin{document}

\title{Scanning tunneling spectroscopy as a probe of multi-Q magnetic states
of itinerant magnets}

\author{Maria N. Gastiasoro}

\affiliation{Niels Bohr Institute, University of Copenhagen, Universitetsparken
5, DK-2100 Copenhagen, Denmark}

\author{Ilya Eremin}

\affiliation{Institut für Theoretische Physik III, Ruhr-Universität Bochum, 44801
Bochum, Germany}

\author{Rafael M. Fernandes}

\affiliation{School of Physics and Astronomy, University of Minnesota, Minneapolis,
Minnesota 55455, USA}

\author{Brian M. Andersen}

\affiliation{Niels Bohr Institute, University of Copenhagen, Universitetsparken
5, DK-2100 Copenhagen, Denmark}
\begin{abstract}
The combination of electronic correlations and Fermi surfaces with
multiple nesting vectors can lead to the appearance of complex multi-\textbf{Q}
magnetic ground states, hosting unusual states such as chiral density-waves
and quantum Hall insulators. 
Distinguishing single-\textbf{Q} and
multi-\textbf{Q} magnetic phases is however a notoriously difficult experimental
problem. Here we propose theoretically
that the local density of states (LDOS) near a magnetic impurity,
whose orientation may be controlled by an external magnetic field,
can be used to map out the detailed magnetic configuration of an itinerant
system and distinguish unambiguously between single-\textbf{Q} and
multi-\textbf{Q} phases. We demonstrate this concept by computing
and contrasting the LDOS near a magnetic impurity embedded in three
different magnetic ground states relevant to iron-based superconductors
-- one single-\textbf{Q} and two double-\textbf{Q} phases. Our results
open a promising avenue to investigate complex magnetic configurations
in itinerant systems via standard scanning tunneling spectroscopy
(STS), without requiring spin-resolved capability. 
\end{abstract}
\maketitle



\begin{figure}[b]
\begin{centering}
\includegraphics[width=0.99\columnwidth]{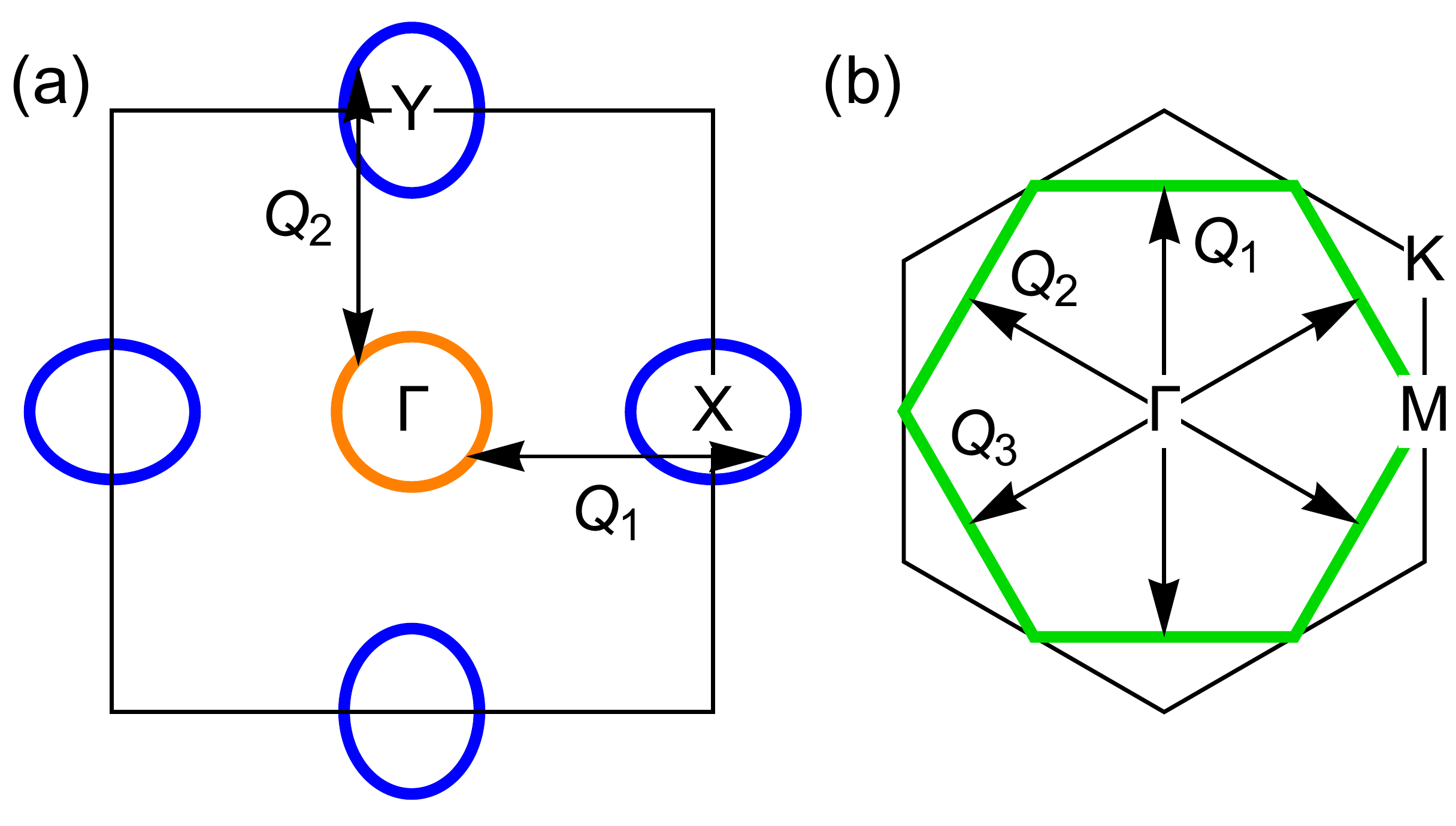} 
\par\end{centering}

\protect\caption{\textbf{Schematic Fermi surfaces}. Illustration of electronic Fermi
surfaces with characteristic nesting vectors with tetragonal (a) and
triangular (b) underlying lattice structures.}

\label{fig:FS} 
\end{figure}

Despite its predominance in localized spin systems \cite{Starykh_review,Balents10},
magnetic frustration is also found in several itinerant systems. While
in most cases magnetic frustration can arise due to the geometry of
the lattice or competing exchange interactions, in purely itinerant
systems it can be manifested as a degeneracy among different nesting-driven
magnetic instabilities with symmetry-related ordering vectors $\mathbf{Q}$
\cite{Hayami_multiQ}. Depending on the symmetry of the lattice and
on the topology of the Fermi surface, different sets of $\mathbf{Q}$
vectors are possible \cite{Venderbos16}. In a square lattice, a compensated
metal with small hole-like and electron-like Fermi pockets, as shown
in Fig.~\ref{fig:FS}(a), has magnetic instabilities at the two nesting
vectors $\mathbf{Q}_{1}=\left(\pi,0\right)$ and $\mathbf{Q}_{2}=\left(0,\pi\right)$,
which are related by a $90^{\circ}$ rotation. The resulting double-\textbf{Q}
phases have been shown to also display charge and vector-chirality
orders \cite{RMF16}. Such a toy model has been widely employed to
study the magnetic properties of iron-based superconductors \cite{eremin2010,RMF_review}
and, more recently, of topological Kondo insulators \cite{Galitski15}.
In the triangular and honeycomb lattices with a hexagonal Fermi surface,
as shown in Fig.~\ref{fig:FS}(b), three nesting vectors related
by $60^{\circ}$ rotations are present, $\mathbf{Q}_{1}=\left(0,\frac{2\pi}{\sqrt{3}}\right)$,
$\mathbf{Q}_{2}=\left(\frac{\pi}{3},-\frac{\pi}{\sqrt{3}}\right)$,
and $\mathbf{Q}_{3}=\left(-\frac{\pi}{3},-\frac{\pi}{\sqrt{3}}\right)$.
Interestingly, the possible triple-\textbf{Q} states display semi-metallic
and quantum Hall insulator behaviors. Such a model has been employed
to study doped cobaltates and graphene doped to the van Hove singularity
point of its band structure \cite{Batista08,Nandkishore12,DHLee12,Thomale12}.
Magnetic instabilities of single- versus triple-$\mathbf{Q}$ phases
of hexagonal Fermi surface were also studied recently in the context
of topological insulators forming magnetic skyrmionic ground state
configurations\cite{kotetes}. In all cases, the presence of repulsive
electronic interactions is essential to stabilize the magnetic instabilities
over other density-wave or superconducting instabilities.

In spite of the interesting properties of multi-\textbf{Q} phases,
unambiguously establishing their existence in a material is a notoriously
difficult experimental problem. For instance, neutron scattering,
which is the prime tool to probe magnetic configurations, is sensitive
not only to the intensity of the order parameters, but also to their
corresponding volume fraction. As such, the neutron scattering signatures
of a multi-\textbf{Q }phase can be nearly indistinguishable from the
signatures of multiple domains of different single-\textbf{Q} phases.
This general issue has been previously highlighted in the literature
in the context of a variety of different materials \cite{Barbara80,Jensen08,Jensen81,Fishman99}.
In some cases, the fact that single-\textbf{Q} or multi-\textbf{Q}
states break additional discrete symmetries of the lattice may facilitate
their experimental distinction. However, in many cases this distortion
may be too small to be resolved experimentally. Several other bulk
probes, such as angle-resolved photo-emission spectroscopy (ARPES),
suffer from similar issues.

\begin{figure}[t]
\begin{centering}
\includegraphics[width=0.9\columnwidth]{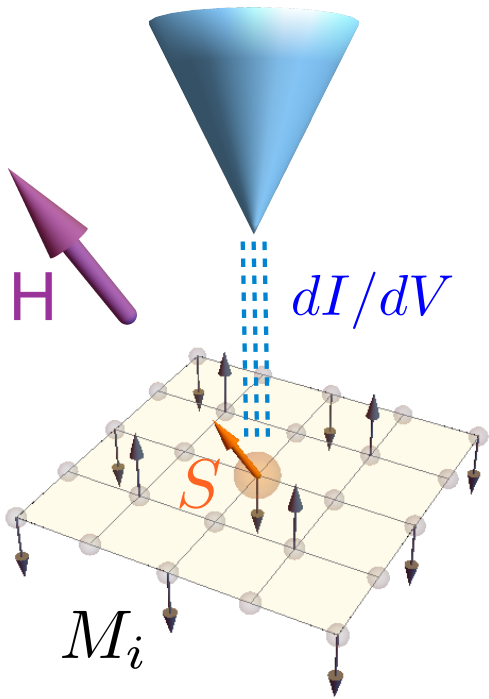} 
\par\end{centering}

\protect\caption{\textbf{Illustration of proposed tunneling experiment}. The total tunneling
conductance measured by an STM tip (blue tip) at the site of a magnetic
impurity moment $\mathbf{S}$ (orange arrow) embedded in a metallic
host of itinerant electrons can be utilized to reveal the detailed
magnetic structure of the host magnetism $\mathbf{M_{i}}$ (black
arrows). The external magnetic field $\mathbf{H}$ is directed along
the purple arrow and determines the direction of the impurity moment.}

\label{fig:STM} 
\end{figure}

This analysis begs the question of whether local probes may be more
appropriate to distinguish single-\textbf{Q} versus multi-\textbf{Q}
phases. Indeed, Mössbauer experiments have been recently employed
to distinguish single-\textbf{Q} versus double-\textbf{Q }phases in
iron-based superconductors \cite{allred16}. In this paper, we propose
theoretically that the magnetic field dependence of the LDOS, which
can be obtained by performing STS measurements near a magnetic impurity,
contains information that allows one to achieve this goal. In a nutshell,
the local changes of the spin density amplitude induced around the
impurity moment, are determined by the specific magnetic ground state
of the host. They cause distinct LDOS signatures since the square
of the local spin amplitude couples to the local charge density. We
explain in detail this general framework and propose experimental
realization for how to manipulate magnetic impurity moments interacting
with the surrounding conduction electrons to pinpoint the nature of
the magnetic ground state of itinerant systems. We illustrate the
proposed setup in Fig.~\ref{fig:STM} and focus for concreteness
on different proposed magnetic ground states relevant for iron-based
superconductors.

Several theoretical calculations \cite{lorenzana08,eremin2010,giovannetti11,Mngeese,kang14,GA2015,christensen2015}
and experimental studies \cite{kim10,avci14,inosov13,hassinger12,wasser15,boehmer15,allred15,mallett1,allred16,mallett2,hassinger15}
have revealed that these materials support not only a single-\textbf{Q}
magnetic stripe (MS) phase, with ordering vectors $\mathbf{Q}_{1}=\left(\pi,0\right)$
or $\mathbf{Q}_{2}=\left(0,\pi\right)$ {[}Fig.~\ref{fig:4}(c){]},
but also two possible types of double-\textbf{Q} phases: a collinear
double-\textbf{Q} phase with non-uniform Fe-magnetization, called
charge-spin density-wave (CSDW) {[}Fig.~\ref{fig:4}(a){]}, and a
coplanar double-\textbf{Q} phase called spin-vortex crystal (SVC)
\cite{RMF16} {[}Fig.~\ref{fig:4}(e){]}. In the former, the staggered
magnetization vectors corresponding to the two $\mathbf{Q}$ vectors
are parallel to each other, whereas in the latter they are perpendicular.

In the proposed STS experiment shown in Fig.~\ref{fig:STM}, a small
fraction of magnetic impurities such as Mn substitutes for Fe ions
in a material which displays one of the possible magnetic configurations
discussed above. On the one hand, the orientation of the itinerant
ordered magnetic moments of the host system is fixed by the residual
spin-orbit and magneto-elastic couplings. On the other hand, the coupling
of the magnetic impurity to the host electronic system is determined
by the Kondo-like interaction between the impurity and the conduction
electrons $J_{K}$ \cite{texier12,leboeuf13}, which in the case of
Mn was found to be small by recent ESR studies \cite{rosa14}. This
property, allied to the insensitivity of the itinerant magnetism to
external magnetic fields \cite{PDai11,enayat14}, implies that magnetic fields
are able to ``unlock'' the magnetic moment of the impurity from
the magnetic order in the lattice, thus allowing for a change of its
orientation with respect to the rigid magnetic structure of the host
system. In the simplest model in which the impurity moment orients
itself parallel to the applied field, we calculate the impurity-modified
LDOS structures as a function of external field direction for all
three distinct magnetic ground states. We find qualitative differences
in the obtained LDOS spectra and demonstrate that scanning tunneling
spectroscopy offers a promising route to unambiguously distinguish
single-\textbf{Q} from multi-\textbf{Q} magnetic phases.

\section*{Microscopic model}

We employ a Hamiltonian relevant to iron-pnictides, which consists
of a five-orbital tight-binding term~\cite{ikeda10} 
\begin{equation}
\mathcal{H}_{0}=\sum_{\mathbf{ij},\mu\nu,\sigma}t_{\mathbf{ij}}^{\mu\nu}c_{\mathbf{i}\mu\sigma}^{\dagger}c_{\mathbf{j}\nu\sigma}-\mu_{0}\sum_{\mathbf{i}\mu\sigma}n_{\mathbf{i}\mu\sigma}.\label{eq:H0}
\end{equation}
Interactions are included through the multi-orbital on-site Hubbard model 
\begin{linenomath}
\begin{align}
\mathcal{H}_{int} & =U\sum_{\mathbf{i},\mu}n_{\mathbf{i}\mu\uparrow}n_{\mathbf{i}\mu\downarrow}+(U'-\frac{J}{2})\sum_{\mathbf{i},\mu<\nu,\sigma\sigma'}n_{\mathbf{i}\mu\sigma}n_{\mathbf{i}\nu\sigma'}\label{eq:Hint}\\
 & \quad-2J\sum_{\mathbf{i},\mu<\nu}\vec{S}_{\mathbf{i}\mu}\cdot\vec{S}_{\mathbf{i}\nu}+J'\sum_{\mathbf{i},\mu<\nu,\sigma}c_{\mathbf{i}\mu\sigma}^{\dagger}c_{\mathbf{i}\mu\bar{\sigma}}^{\dagger}c_{\mathbf{i}\nu\bar{\sigma}}c_{\mathbf{i}\nu\sigma},\nonumber 
\end{align}
\end{linenomath}
where $\mu,\nu$ are orbital indices, ${\mathbf{i}}$ denotes lattice
sites, and $\sigma$ is the spin. The interaction includes intraorbital
(interorbital) repulsion $U$ ($U'$), the Hund's coupling $J$, and
the pair hopping term $J'$. Following previous studies~\cite{GA2015}
we assume spin and orbital rotation invariance, implying $U'=U-2J$
and $J'=J$, and fix $J=U/4$. As shown previously in Ref. \onlinecite{GA2015},
the Hamiltonian $\mathcal{H}=\mathcal{H}_{0}+\mathcal{H}_{int}$ supports
all three magnetic ground states depending on interaction parameters
and electron filling (details in the Supplementary Material (SM)).
The resulting reconstructed Fermi surfaces in the magnetic phases
are remarkably similar, and do not constitute a good probe of the
preferred ordered phase (see also SM).

\begin{figure*}[t!]
\begin{centering}
\includegraphics[width=0.99\textwidth]{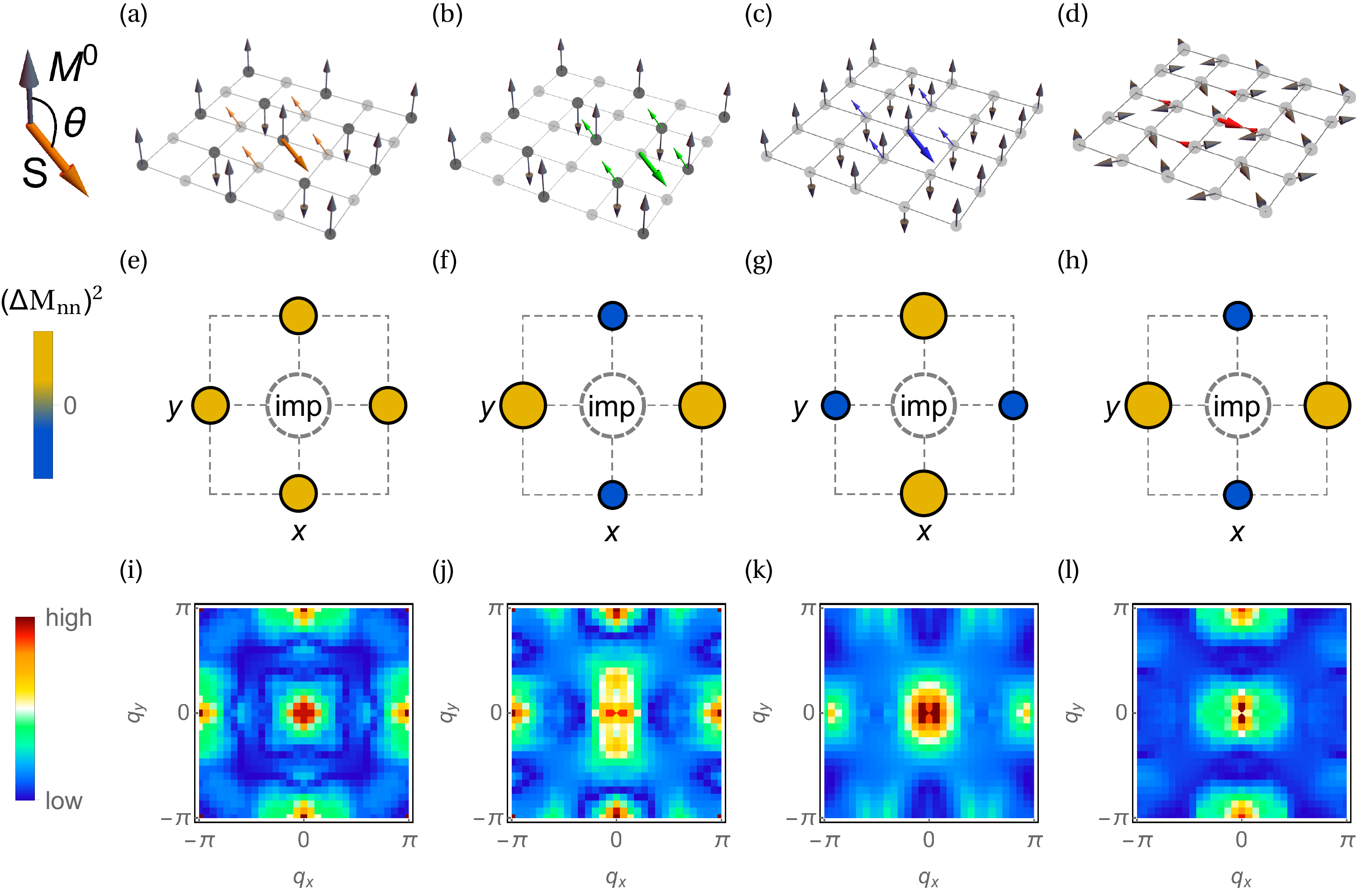} 
\par\end{centering}

\protect\caption{\textbf{Connection between impurity-modified LDOS and the magnetic
ground state}. Magnetic impurity (colored thick arrows) with $\theta=3\pi/4$
orientation in the CSDW state (either in a magnetic site (a) or in
a non-magnetic site (b)), in the MS state, and in the SVC state (d).
The colored (black) arrows represent the induced (impurity-free) magnetization
$\mathbf{M}_{\mathbf{nn}}^{\mathrm{ind}}$ at the nn sites ($\mathbf{M}_{\mathbf{i}}^{0}$).
(e)-(h) Sketches of the total moment amplitude deviation $(\Delta M)^{2}$
compared to the impurity-free case at the four nn sites ($(\Delta M_{\mathbf{nn}})^{2}\equiv[(\mathbf{M}_{\mathbf{nn}}\cdot\mathbf{\hat{l}})^{2}-(\mathbf{M}_{\mathbf{nn}}^{0}\cdot\mathbf{\hat{l}})^{2}]\mathbf{\hat{l}}$)
in the previous four cases for (e)-(g) $\mathbf{\hat{l}}=\mathbf{\hat{z}}$
and (h) $\mathbf{\hat{l}}=\mathbf{\hat{x}}$ projection. Yellow (blue)
denotes increased (decreased) amplitude. (i)-(l) The Fourier transformed
LDOS $N(\mathbf{q},\theta=3\pi/4)$ of the (a)-(d) impurity configurations,
respectively.}

\label{fig:3} 
\end{figure*}

The magnetic impurity moment is included by the following term in
the Hamiltonian 
\begin{equation}
\mathcal{H}_{S}=\frac{J_{K}}{2}\sum_{\mu}\mathbf{S}_{\mathbf{i^{*}}}\cdot\left(c_{\mathbf{i^{*}}\mu\sigma}^{\dagger}\boldsymbol{\sigma}_{\sigma\sigma'}c_{\mathbf{i^{*}}\mu\sigma'}\right)-g\mu_{B}\mathbf{S}_{\mathbf{i^{*}}}\cdot\mathbf{H}.\label{eq_Kondo}
\end{equation}
The first term corresponds to the Kondo-like exchange coupling between
the impurity moment $\mathbf{S}_{\mathbf{i^{*}}}$, located at site
$\mathbf{i}^{*}$, and the spin of the itinerant electrons, which
we denote hereafter by $\mathbf{M}_{\mathbf{i}}^{0}\equiv\sum_{\mu}c_{\mathbf{i}\mu\sigma}^{\dagger}\boldsymbol{\sigma}_{\sigma\sigma'}c_{\mathbf{i}\mu\sigma'}$.
The second term corresponds to the Zeeman coupling between the impurity
moment and the external magnetic field $\mathbf{H}$. In principle,
one would also need to include the Zeeman coupling between the itinerant
moments and $\mathbf{H}$. However, as shown experimentally e.g. in Refs.
\onlinecite{PDai11,enayat14}, the itinerant magnetization is insensitive to magnetic
fields of the order of $10$ T, and therefore this term can be safely
neglected. The main question is whether the external field $\mathbf{H}$
is capable of rotating the impurity moment, which is also coupled
to the itinerant electrons via $J_{K}$. To answer this question,
we rely on ESR experiments that measured $J_{K}$ for Mn-doped iron
pnictides \cite{rosa14}. The reported value $J_{K}\approx0.8$ meV
is very small, suggesting that magnetic fields of the order of $10$
T (achievable in STM setups) can unlock the impurity moment from the
itinerant magnetic configuration. Importantly, the fact that the Mn
impurities act as local magnetic moments and do not add charge carriers
into the system is supported by the NMR data in Refs. \onlinecite{inosov13,leboeuf13}.

Therefore, we proceed by fixing the direction of $\mathbf{S}_{\mathbf{i^{*}}}$
to be parallel to $\mathbf{H}$, and introduce the polar angle $\theta$
between the magnetization of the itinerant electrons in the impurity-free
system $\mathbf{M}_{\mathbf{i^{*}}}^{0}$ and the impurity moment
$\mathbf{S}_{\mathbf{i^{*}}}$ at that site, $\cos\theta\propto\mathbf{S}_{\mathbf{i^{*}}}\cdot\mathbf{M}_{\mathbf{i^{*}}}^{0}$,
as illustrated in Fig.~\ref{fig:3}(a). We use a magnetic impurity moment with $J_K S=0.1$meV, a value that is not important since our goal is to focus on the
symmetry changes of the LDOS as the impurity moment rotates, i.e.
as a function of the orientation $\theta$. Thus, we solve Eqs. (\ref{eq:H0}),
(\ref{eq:Hint}), and (\ref{eq_Kondo}) and calculate the LDOS $N(\mathbf{i},\theta)$
at each lattice site according to 
\begin{equation}
N(\mathbf{i},\theta)=-\frac{1}{\pi}\mbox{Im}\sum_{n\mu\sigma}\frac{u_{\mu\sigma}^{n}(\mathbf{i},\theta)u_{\mu\sigma}^{n}(\mathbf{i},\theta)}{\omega-E_{n}(\theta)+i\eta}.
\end{equation}
Here $u_{\mu\sigma}^{n}(\mathbf{i},\theta)$ are the matrix elements
of the unitary transformation from orbital $\mu$ to eigenstate $n$.
For further computational details, and spectral studies of the homogenous
magnetic phases and the effects from non-magnetic disorder, we refer
to the SM. The results are presented in Fig.~\ref{fig:3} and discussed
in details below.

\section*{Results}

\emph{The CSDW state.} We start by discussing the results in the CSDW
state. As illustrated in Figs.~\ref{fig:3}(a) and \ref{fig:3}(b),
in this collinear non-uniform double-\textbf{Q} magnetic phase the
even sites of the square lattice are non-magnetic, whereas the odd
sites display a Neel-like magnetic configuration with spins parallel
to the $z$ direction. As a result, there are four inequivalent sites
to place an impurity, two magnetic ($\mathbf{M}_{i^{*}}^{0}\neq0$)
and two non-magnetic ($\mathbf{M}_{\mathbf{i^{*}}}^{0}=0$) ones.
Let us start discussing the modified magnetization around the impurity,
which will be necessary to understand the resulting spectral signatures.
The black arrows display the magnetization of the conduction electrons
of the impurity-free system $\mathbf{M}_{\mathbf{i}}^{0}$, and the
thick colored arrow indicates the impurity moment $\mathbf{S}_{\mathbf{i^{*}}}$.
The induced spin density on the nearest neighbor (nn) sites $\mathbf{M}_{\mathbf{nn}}^{\mathrm{ind}}$
is illustrated by the four arrows of the same color. The sum of the
induced and impurity-free magnetizations yields the new total magnetization
in the presence of the impurity, $\mathbf{M}_{\mathbf{i}}=\mathbf{M}_{\mathbf{i}}^{0}+\mathbf{M}_{\mathbf{i}}^{\mathrm{ind}}$.
In our calculation, the induced spin density actually involves a larger
number of sites surrounding the impurity moment, but for the symmetry
arguments used below, it is sufficient to focus on the nn sites, where
the effect is the largest.

We sketch in Fig.~\ref{fig:3}(e) the change in magnetic moment (projected along the $\mathbf{\hat{l}}=\mathbf{\hat{z}}$ axis) at the four nn
sites induced by an impurity oriented along $\theta=3\pi/4$, $(\Delta M_{\mathbf{nn}})^{2}\equiv(\mathbf{M}_{\mathbf{nn}}\cdot\mathbf{\hat{l}})^{2}-(\mathbf{M}_{\mathbf{nn}}^{0}\cdot\mathbf{\hat{l}})^{2}$.
Clearly, the magnetization amplitude increases equally at all four
nn sites, i.e. $|\mathbf{M}_{\mathbf{r}_{1}}\cdot\mathbf{\hat{l}}|=|\mathbf{M}_{\mathbf{r}_{2}}\cdot\mathbf{\hat{l}}|$,
where $\mathbf{r}_{1}=\mathbf{i^{*}}+\mathbf{\hat{x}}$ and $\mathbf{r}_{2}=\mathbf{i^{*}}+\mathbf{\hat{y}}$
denote the two types of nn sites. If the same impurity moment is placed
at a non-magnetic site, however, as illustrated in Fig.~\ref{fig:3}(b),
the total projected moments on $\mathbf{r}_{1}$ and $\mathbf{r}_{2}$
become unequal. In particular, while the projected moment is reduced
at the two sites along the $y$ axis, it is enhanced at the sites
along the $x$ axis. This antagonistic change is sketched in Fig.~\ref{fig:3}(f).
Consequently, the initial tetragonal symmetry of the magnetization
is locally broken by the impurity, with $|\mathbf{M}_{\mathbf{r}_{1}}\cdot\mathbf{\hat{z}}|\neq|\mathbf{M}_{\mathbf{r}_{2}}\cdot\mathbf{\hat{z}}|$.
For a general angle $\theta$ between the impurity moment and the
itinerant magnetization, this symmetry breaking is given by the following
expression,
\begin{linenomath}
\begin{align}
|\mathbf{M}_{\mathbf{r}_{1}}|^{2}-|\mathbf{M}_{\mathbf{r}_{2}}|^{2} & =\left(|\mathbf{M}_{\mathbf{r}_{1}}^{\mathrm{ind}}|^{2}-|\mathbf{M}_{\mathbf{r}_{2}}^{\mathrm{ind}}|^{2}\right)\label{eq:diff}\\\nonumber
 & -2\left(\mathbf{M}_{\mathbf{nn}}^{0}\cdot\mathbf{\hat{z}}\right)\left(|\mathbf{M}_{\mathbf{r}_{1}}^{\mathrm{ind}}|+|\mathbf{M}_{\mathbf{r}_{2}}^{\mathrm{ind}}|\right)\cos\theta.
\end{align}
\end{linenomath}
To make the argument more transparent, we assumed an anti-parallel
orientation of $\mathbf{M}_{\mathbf{r}_{i}}^{\mathrm{ind}}$ with
respect to the impurity moment, which is strictly correct for an arbitrary
$\theta$ only in the paramagnetic case, but it remains a reasonable
approximation in our case. The full self-consistent result beyond
this assumption is presented below.

To understand these results, we note that impurities at the magnetic
sites of the CSDW phase {[}Fig.~\ref{fig:3}(a){]} are subject to
two constrains. First, the impurity-free SDW has nodes at the nn sites
($\mathbf{M}_{\mathbf{nn}}^{0}\cdot\mathbf{\hat{z}}=0$), and second,
the symmetry of this site requires the induced moments to be the same,
$\mathbf{M}_{\mathbf{r}_{1}}^{\mathrm{ind}}=\mathbf{M}_{\mathbf{r}_{2}}^{\mathrm{ind}}$.
Hence, $|\mathbf{M}_{\mathbf{r}_{1}}|^{2}=|\mathbf{M}_{\mathbf{r}_{2}}|^{2}$
and tetragonal symmetry is preserved for any orientation of the impurity
moment. By contrast, for impurity moments at non-magnetic sites, the
difference between the total nn amplitudes has contributions from
both terms in Eq.~\eqref{eq:diff}. Therefore, the symmetry-breaking
expression is generally non-zero, and exhibits a cosine-like directional
dependence.

Having established the basic effect of an impurity moment on the surrounding
itinerant magnetic structure, we now study its consequences on the
local spectral features. The total LDOS measures the spectral composition
of the charge density, which is coupled by symmetry only to the magnetization
density squared. Therefore, any change in the amplitude of the spin
density will have an impact on the LDOS. For instance, a $C_{4}$
($C{}_{2}$) symmetric magnetic structure will generally present $C{}_{4}$
($C{}_{2}$) symmetric LDOS signatures (here $C_{4}$ and $C_{2}$
denote tetragonal and orthorhombic symmetries). Fig.~\ref{fig:3}(i)
shows the Fourier transformed LDOS of the case displayed in Fig.~\ref{fig:3}(a),
$N(\mathbf{q},3\pi/4)$. Clearly, the $C_{4}$ symmetry of the pristine
CSDW state is preserved around the impurity, in agreement with the
magnetic moment structure sketched in Fig.~\ref{fig:3}(e) and the
result $|\mathbf{M}_{\mathbf{r}_{1}}|^{2}=|\mathbf{M}_{\mathbf{r}_{2}}|^{2}$
given by Eq.~\eqref{eq:diff}. The spectral symmetry is in fact tetragonal
for all possible orientations of the impurity in the CSDW phase, as
long as the impurity moment is placed at a magnetic site. If the same
impurity moment is placed at a non-magnetic site, however, as illustrated
in Fig.~\ref{fig:3}(b), $N(\mathbf{q},3\pi/4)$ becomes $C_{2}$
symmetric as seen in Fig.~\ref{fig:3}(j). This is simply a consequence
of the $C{}_{2}$ symmetric spin structure induced by the impurity
discussed earlier and sketched in Fig.~\ref{fig:3}(f).

\begin{figure*}[t]
\begin{centering}
\includegraphics[width=0.99\textwidth]{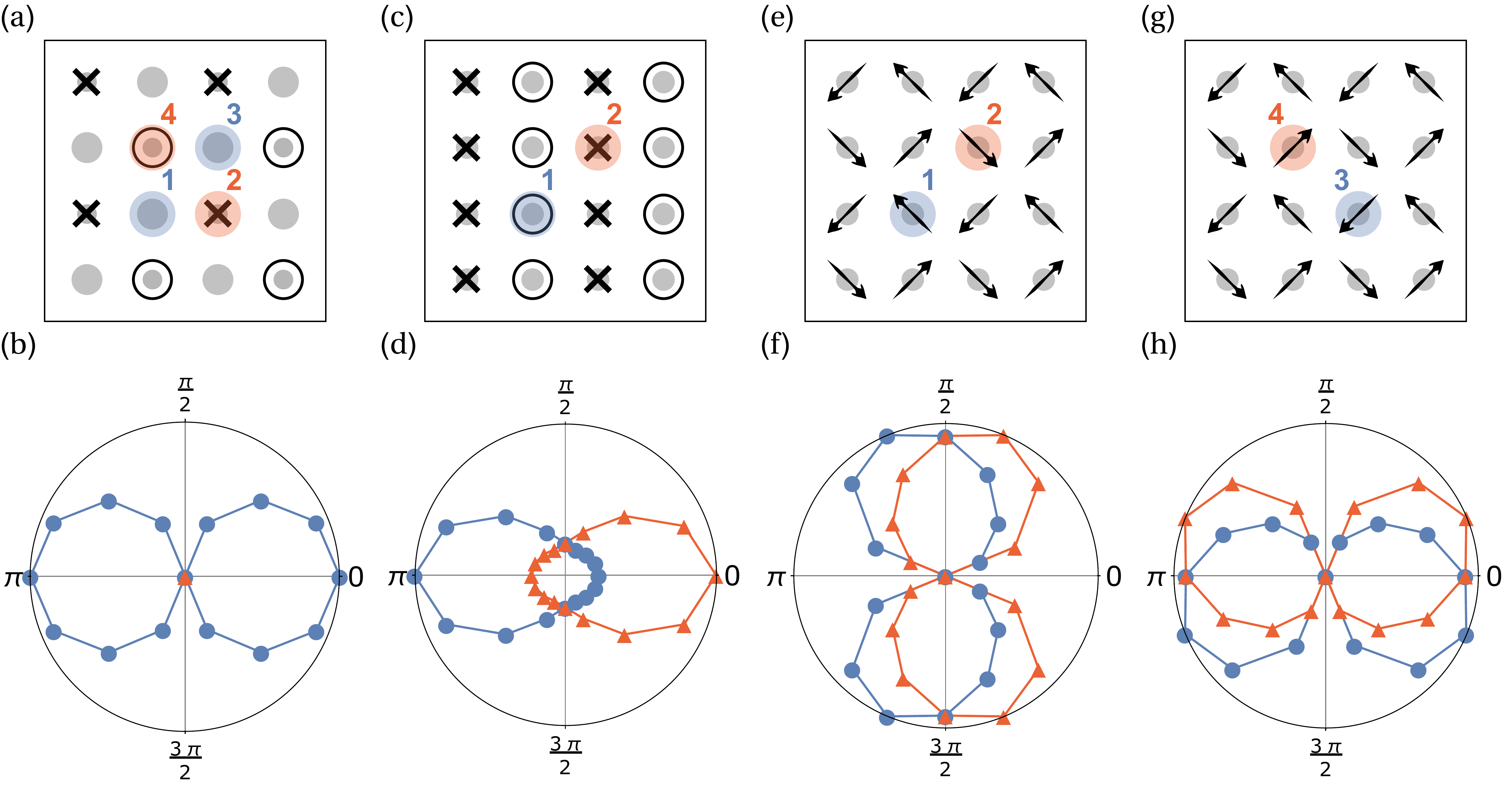} 
\par\end{centering}

\protect\caption{\textbf{Angular dependence of $C_{4}$-symmetry breaking in the LDOS}.
Spectral symmetry breaking parameter $\delta_{C_{2}}(\theta)$ {[}see
Eq.~\eqref{eq:delta}{]} as a function of the orientation of the
magnetic impurity $\theta$ in the inequivalent sites of the magnetic
states. $\theta$ represents the polar angle in the (a)-(b) CSDW and
(c)-(d) MS states, and (e)-(h) the azimuthal angle in the SVC state.
$\theta=3\pi/4$ examples are explicitly shown in Fig.~\ref{fig:3}.
(b),(d), (f) and (h) are polar plots of the average $\delta_{C_{2}}(\theta)$
as radius in a given direction $\theta$ in the $xz$ ($xy$) plane
for all relevant inequivalent sites of the CSDW and MS (CSV) states.}

\label{fig:4} 
\end{figure*}

In order to quantify the evolution of the spectral symmetry of the LDOS
as the impurity rotates, and compare it to Eq.~\eqref{eq:diff},
we introduce the anisotropy parameter 
\begin{equation}
\delta_{C_{2}}(\theta)=\sum_{\mathbf{i}}\frac{|N(\mathbf{i},\theta)-N(R\{\mathbf{i}\},\theta)|}{N(\mathbf{i},\theta)+N(R\{\mathbf{i}\},\theta)},\label{eq:delta}
\end{equation}
which measures the breaking of the $C_{4}$ symmetry for a given $\theta$.
Here $R$ denotes a $\pi/2$ rotation operation. Figure~\ref{fig:4}(b)
shows the evolution of $\delta_{C_{2}}$ as a function of $\theta$
from the calculated LDOS at all four inequivalent sites in the CSDW
state, specified in Fig.~\ref{fig:4}(a). As seen, tetragonal symmetry
is indeed preserved for all $\theta$ at the magnetic sites (2 and
4), but broken in a cosine-like fashion at the non-magnetic sites
(1 and 3), in agreement with the above discussion.

\emph{The MS state.} The single-$\mathbf{Q}$ MS phase corresponds
to the case in which only one of the two possible $\mathbf{Q}$ vectors
is selected. The corresponding magnetic configuration, shown in Fig.~\ref{fig:3}(c),
breaks the $C_{4}$ symmetry down to $C_{2}$, in contrast to the
double-\textbf{Q} magnetic configurations, which preserve tetragonal
symmetry. We now consider the effects of an impurity moment with $\theta=3\pi/4$
orientation with respect to the SDW magnetization. The inequivalent
change in the spin projection along the $z$ axis of the four nn sites
is illustrated in Fig.~\ref{fig:3}(g), which accounts for the symmetry
breaking in the corresponding $N(\mathbf{q},3\pi/4)$ shown in Fig.~\ref{fig:3}(k).
The evolution of $\delta_{C_{2}}(\theta)$ for the two inequivalent
sites in this state can be seen in Fig.~\ref{fig:4}(d). Clearly,
both sites (specified in Fig.~\ref{fig:4}(c)) give rise to distinct
angular evolutions which break $C{}_{2}$ symmetry. This is in contrast
to what was found in the CSDW state, where half of the sites exhibit
a $C{}_{2}$-symmetric cosine-like evolution of $\delta_{C_{2}}(\theta)$,
and the other half preserve $C_{4}$ symmetry for all impurity orientations,
$\delta_{C_{2}}(\theta)=0$.

\emph{The SVC state.} Finally we discuss the case of a magnetic impurity
in the coplanar SVC phase. In this double-\textbf{Q} magnetic state,
the even sites of the square lattice display a Neel-like order that
is perpendicular to the Neel-like order displayed by the odd sites.
As a result, there are four inequivalent sites. One of them is shown
in Fig.~\ref{fig:3}(d), with the calculated LDOS displayed in Fig.~\ref{fig:3}(l).
Again, the $C_{2}$ symmetric spectral features can be connected to
the different moment amplitudes at the nn sites, as illustrated in
Fig.~\ref{fig:3}(h) for projections along the $x$ axis. The angular
evolution of $\delta_{C_{2}}(\theta)$ in the four inequivalent sites
(in the $xy$ plane for this state) is shown in Fig.~\ref{fig:4}(f,h).
Contrary to what was found in the CSDW phase, all inequivalent sites
show a $\delta_{C_{2}}(\theta)$ spectral angular dependence with
broken $C{}_{2}$ symmetry. Moreover, the angular dependences appear
identical but shifted by $n\pi/2$ for all four inequivalent sites
($n=1,2,3$), which is a consequence of the uniform moment amplitude
$|\mathbf{M}_{\mathbf{i}}^{0}|=M^{0}$ in the SVC state.

\section*{Discussion}

The results presented in the previous section open different routes
to distinguish between single-\textbf{Q }and multi-\textbf{Q }magnetic
ground states via STS experiments. The
most direct way would be to extract the LDOS anisotropy parameter
$\delta_{C_{2}}$ as function of the angle $\theta$ between the applied
magnetic field and the magnetization. Note that a recent STS experiment
on the iron pnictide NaFeAs extracted precisely this anisotropy parameter
(for zero magnetic field) \cite{Pasupathy14}. As shown in Fig. \ref{fig:4},
the function $\delta_{C_{2}}(\theta)$ behaves qualitatively different
for each of the three magnetic ground states: 
\begin{itemize}
\item In the CSDW state, experiments would only observe a single anisotropy
parameter curve $\delta_{C_{2}}(\theta)$, which displays a cosine-like
$C{}_{2}$ symmetric shape {[}Fig.~\ref{fig:4}(b){]}. This curve
corresponds to a magnetic impurity placed in a magnetic site, since
impurities located at non-magnetic sites exhibit fully $C_{4}$-symmetric
LDOS.
\item In the MS state, experiments would observe two anisotropy parameter
curves $\delta_{C_{2}}(\theta)$ {[}Fig.~\ref{fig:4}(d){]}. These
two curves, related to the two inequivalent site positions in the
magnetic ground state, are simply related by a $\theta=\pi$ shift.
In addition, tetragonal symmetry is broken for \emph{all} impurity
orientations, i.e. $\delta_{C_{2}}(\theta)\neq0$ for all $\theta$
values.
\item In the SVC state, experiments would observe four anisotropy parameter
curves $\delta_{C_{2}}(\theta)$ {[}Fig.~\ref{fig:4}(f,h){]}. They
also correspond to the four inequivalent site positions of the SVC
state, and are related to each other by successive shifts of $\theta=\pi/2$. 
\end{itemize}
While a continuous rotation of the magnetic field may be experimentally
challenging, our results can also be useful in cases where the direction
of the field can only be changed from ``positive'' to ``negative.''
Assume, for example, that only the angles $\theta=0$ and $\theta=\pi$
are experimentally accessible. Then, as can be inferred from panels
(b), (d) and (f) in Fig.~\ref{fig:4}, one may still use the measured
LDOS to deduce the true ground state. Alternatively, the Fourier transformation
of the LDOS for two fixed angles, can also be used to constrain the
possible ground states -- see panels (i)-(l) in Fig. \ref{fig:3}.

The proposed experiment advocated in this paper relies on the fact
that an impurity moment can be controlled by an external applied field
which, however, does not significantly affect the itinerant magnetic
order. In addition, it would be desirable if this impurity did not
introduce additional charge carriers in the system. In the case of
the iron-pnictides, Mn impurities are natural candidates, since they
form local moments weakly coupled to the itinerant system \cite{rosa14,inosov13,leboeuf13},
which is itself robust against moderate magnetic fields \cite{PDai11}.
It will be interesting to extend our calculations to multi-\textbf{Q}
magnetic phases in triangular lattices, where exotic triple-\textbf{Q}
phases can appear. Our work provides a promising avenue for future
tunneling spectroscopy to directly distinguish between nearly degenerate
but symmetry-distinct magnetic ground states of itinerant magnetic
systems. 
\begin{acknowledgments}
M.N.G. and B.M.A, acknowledge support from Lundbeckfond fellowship (grant A9318). R.M.F is supported by the Office
of Basic Energy Sciences, U.S. Department of Energy, under award DE-SC0012336. \end{acknowledgments}


\pagebreak
\widetext
\begin{center}
\textbf{\large Supplemental Materials: "Scanning tunneling spectroscopy as a probe of multi-Q magnetic states of itinerant magnets"}
\end{center}

\setcounter{equation}{0}
\setcounter{figure}{0}
\setcounter{table}{0}
\setcounter{page}{1}
\makeatletter
\renewcommand{\theequation}{S\arabic{equation}}
\renewcommand{\thefigure}{S\arabic{figure}}
\renewcommand{\bibnumfmt}[1]{[S#1]}
\renewcommand{\citenumfont}[1]{S#1}

Here, we provide computational details and a band structure description of the paramagnetic and magnetic states. 
We also include a brief spectroscopic study of non-magnetic impurities in the three magnetic states, where we discuss both spin-summed and spin-resolved features.

\section{Computational details}

A mean field decoupling in the spin and charge channels leads to the following total Hamiltonian
\begin{eqnarray}
\mathcal{H}^{MF}=\sum_{\mathbf{ij}\mu\nu}
\begin{pmatrix}
 \hat{c}_{\mathbf i\mu\uparrow}^{\dagger} & \hat{c}_{\mathbf i\mu\downarrow}^{\dagger}
\end{pmatrix}
\begin{pmatrix}
\varphi_{\mathbf{ij}\uparrow}^{\mu\nu} & \omega_{\mathbf{ii}\uparrow}^{\mu\nu}  \\
\omega_{\mathbf{ii}\downarrow}^{\mu\nu} & \varphi_{\mathbf{ij}\downarrow}^{\mu\nu}
\end{pmatrix}
\begin{pmatrix}
 \hat{c}_{\mathbf{j}\nu\uparrow}\\\hat{c}_{\mathbf{j}\nu\downarrow}
\end{pmatrix},
\label{Seq:H}
\end{eqnarray}
where $c_{\mathbf i \mu\sigma}^{\dagger}$ creates an electron at site $\mathbf i$ with spin $\sigma$ in orbital state $\mu$.
$\varphi_{\mathbf{ij}\sigma}^{\mu\nu}$ and $\omega_{\mathbf{ii}\sigma}^{\mu\nu}$ are given by
\begin{align}
  \varphi_{\mathbf{ij}\sigma}^{\mu\nu}&=t_{\mathbf{ij}}^{\mu\nu}+\delta_{\mu\nu}[-\mu_{0}+(\Omega^0_{\mu}+\Omega^z_{\mu})\delta_{\mathbf{ii^*}}\delta_{\mu\nu}+U\langle\hat{n}_{\mathbf i\mu\overline{\sigma}}\rangle+U'\langle\hat{n}_{\mathbf i\nu\overline{\sigma}}\rangle+(U'-J) \langle\hat{n}_{\mathbf i\nu\sigma}\rangle]\\\nonumber
  &\quad-\bar\delta_{\mu\nu}[(U'-J)\langle \hat{c}_{\mathbf i\nu\sigma}^{\dagger}\hat{c}_{\mathbf i\mu\sigma} \rangle +J \langle \hat{c}_{\mathbf i\nu\overline{\sigma}}^{\dagger}\hat{c}_{\mathbf i\mu\overline{\sigma}} \rangle+J'\langle \hat{c}_{\mathbf i\mu\overline{\sigma}}^{\dagger}\hat{c}_{\mathbf i\nu\overline{\sigma}} \rangle], \\
   \omega_{\mathbf{ii}\sigma}^{\mu\nu}&=\delta_{\mu\nu}[\Omega^x_{\mu}\delta_{\mathbf{ii^*}}\delta_{\mu\nu}-U \langle \hat{c}_{\mathbf i\mu\overline{\sigma}}^{\dagger}\hat{c}_{\mathbf i\mu\sigma} \rangle - J\langle \hat{c}_{\mathbf i\nu\overline{\sigma}}^{\dagger}\hat{c}_{\mathbf i\nu\sigma} \rangle] -\bar\delta_{\mu\nu}[U'\langle \hat{c}_{\mathbf i\nu\overline{\sigma}}^{\dagger}\hat{c}_{\mathbf i\mu\sigma} \rangle+J'\langle \hat{c}_{\mathbf i\mu\overline{\sigma}}^{\dagger}\hat{c}_{\mathbf i\nu\sigma} \rangle],
\end{align}
with $\bar\delta_{\mu\nu}=1-\delta_{\mu\nu}$. 
The terms $\Omega^z_{\mu}=\sigma J_K \mathbf S_{\mu}\cdot \hat{z}$ and $\Omega^x_{\mu}= J_K \mathbf S_{\mu}\cdot \hat{x}$ include out-of-plane and in-plane components of a magnetic impurity at site $\mathbf{i^*}$, respectively.
A non-magnetic scatterer can be also introduced at the same site by the $\Omega^0_{\mu}=V_{\mu}$ term. 
In the main text this last term was set to zero, but for a brief non-magnetic impurity study the reader is refered to the next section.
We diagonalize Eq.\eqref{Seq:H} on $30 \times 30$ lattices by a unitary transformation  
$\hat{c}_{\mathbf i\mu\sigma}= \sum_n u_{\mu\sigma}^n (\mathbf i)\hat{\gamma}_{n}$, and the following unrestricted fields are obtained self-consistently 
 \begin{align}
 \label{Seq:fields}
  \langle \hat{c}_{\mathbf i\mu\sigma}^{\dagger} \hat{c}_{\mathbf j\nu\sigma}\rangle&=\sum_{n} u_{\mu\sigma}^{n*}(\mathbf i)u_{\nu\sigma}^{n}(\mathbf j) f(E_n), \\\nonumber
  \langle \hat{c}_{\mathbf i\mu\sigma}^{\dagger} \hat{c}_{\mathbf i\nu\bar\sigma}\rangle&=\sum_{n} u_{\mu\sigma}^{n*}(\mathbf i) u_{\nu\bar\sigma}^{n}(\mathbf i) f(E_n),   
 \end{align}
for all sites $\mathbf i, \mathbf j$ and orbital combinations $\mu,\nu$. Here $E_n$ denote the eigenvalues, and $f$ is the Fermi function.
The magnetization density $\mathbf M_{\mathbf i}=\sum_{\mu\sigma\sigma'} \langle \hat{c}_{\mathbf i\mu\sigma}^{\dagger}  \mathbold\sigma_{\sigma\sigma'} \hat{c}_{\mathbf i\mu\sigma'}\rangle$ ($\mu_B=1$, $g=2$) and charge density $n_{\mathbf i}=\sum_{\mu\sigma}\langle \hat{c}_{\mathbf i\mu\sigma}^{\dagger} \hat{c}_{\mathbf i\mu\sigma}\rangle$ are obtained from the self-consistent fields in Eq.~\eqref{Seq:fields}.

\section{Homogeneous band structure}

\begin{figure}[t]
\begin{center}
\includegraphics[width=0.5\columnwidth]{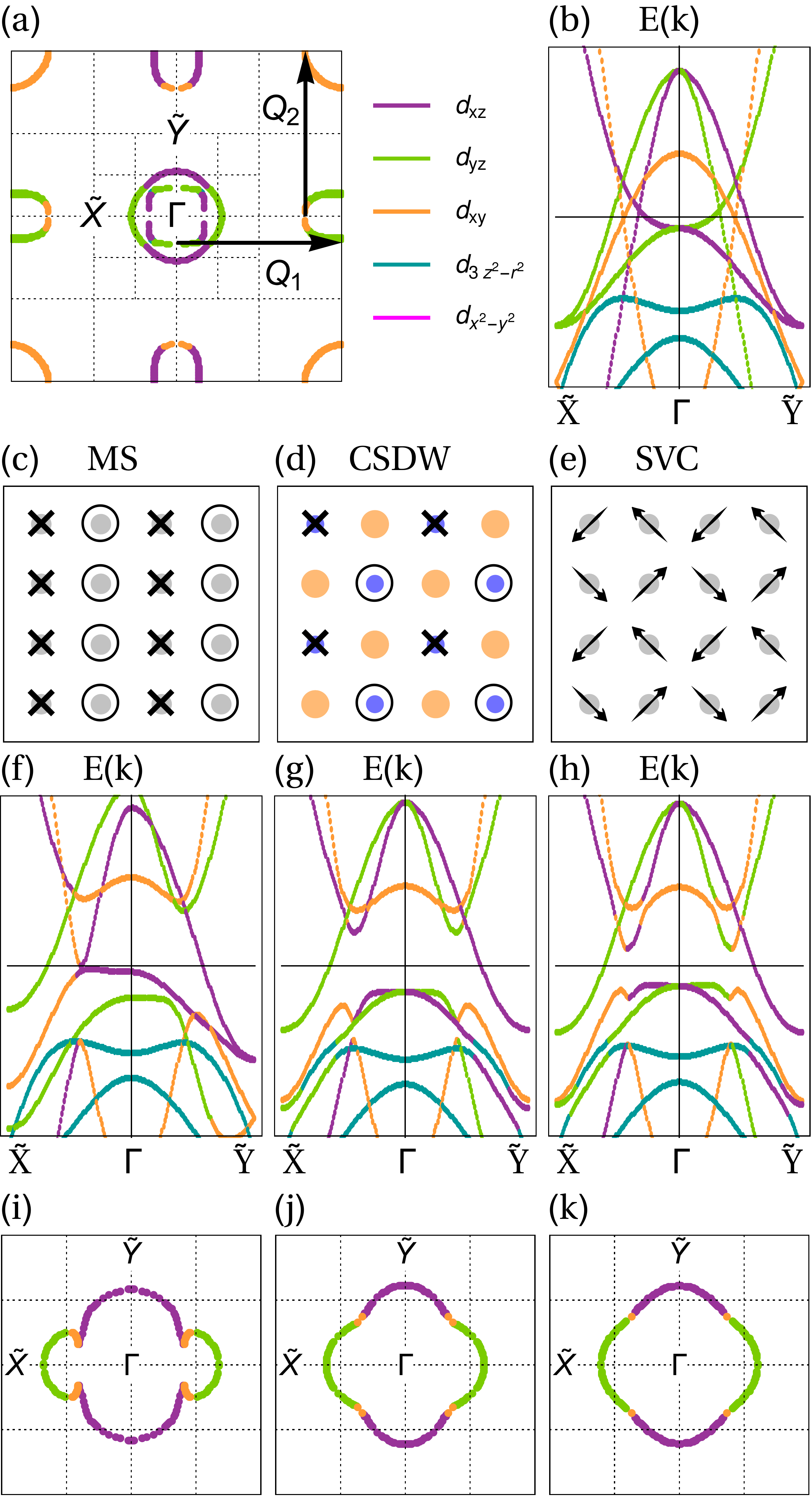}
\end{center}
\caption{(a) Fermi surface of the normal state and (b) its band dispersion $E(k)$ along high symmetry lines in the folded BZ (FBZ) obtained by folding the original BZ by $\mathbf{Q_1}$ and $\mathbf{Q_2}$.  The colors represent the main orbital content, specified in the legend.
(c-e) The three SDW states MS (c), CSDW (d) with out-of-plane oriented moments and $(\pi,\pi)$ charge order with orange (blue) indicating high (low) electron density $n$, and SVC (e). The band dispersions (f)-(h) and Fermi surfaces in the FBZ (i)-(k) for the three respective SDW phases in (c-e).}
\label{fig:1}
\end{figure}

For the band structure relevant to the iron pnictides, three hole-like and two electron-like bands cross the Fermi level, with two degenerate nesting vectors $\mathbf{Q_1}=(\pi,0)$ and $\mathbf{Q_2}=(0,\pi)$, as shown in Fig.~\ref{fig:1}(a,b).
The corresponding RPA spin susceptibility is peaked at $\mathbf{Q_1}$ and $\mathbf{Q_2}$, and in general the spin ordered state will be a combination of both ordering vectors, $\mathbf{M}(\mathbf{r})=\sum_{l=1,2} \mathbf{M}_l \exp(i \mathbf{Q}_l \cdot \mathbf{r})$.~\cite{Slorenzana08,Seremin2010,Sgiovannetti11} 
The spin structures of the three distinct magnetic states are shown in Figs.~\ref{fig:1}(c)-(e). 
For these systems, the prevalent magnetic phase is the MS state with in-plane moments along the antiferromagnetic ordering vector. 
Interestingly, the new experimentally observed magnetic phase exhibits a spin reorientation from in-plane to out-of-plane, with moments along the $c$ axis.~\cite{Swasser15} 
The relevant spin structures for this new phase are thus collinear single-$\mathbf{Q}$ MS and double-$\mathbf{Q}$ CSDW states with out-of-plane moments, which we consider in this work (Figs.~\ref{fig:1}(c)-(d)).  

In order to compare the resulting electronic properties of the three SDW states at an equal footing, we fix the temperature to $\kappa_B T=10$ meV, the electron filling $n=5.88$, and the interaction parameters $U=0.95$ eV. 
In this region of parameter space, the CSDW state is the global minimum.~\cite{SGA2015}
The other two magnetic states are local minima, which may be stabilized self-consistently by applying restrictions to the fields.
In this way all three magnetic states are generated from the same normal state [Figs.~\ref{fig:1}(a)-(b)].
Figures~\ref{fig:1}(f)-(k) show the reconstructed Fermi surfaces and band structures of the three different SDW states along high symmetry lines in the folded Brillouin zone (FBZ) ($-\pi/2<k_{x},k_y<\pi/2$).
In the single-$\mathbf{Q}$ MS state only SDW gaps at momenta connected by the ordering vector $\mathbf{Q_1}$ open, leaving the direction parallel to the stripes metallic, and thus resulting in a $C_2$ symmetric band dispersion and Fermi surface shown in Figs.~\ref{fig:1}(f) and \ref{fig:1}(i), respectively.
In the double-$\mathbf{Q}$ states, the gaps open at momenta connected by both $\mathbf{Q_1}$ and $\mathbf{Q_2}$, resulting in two very similar band reconstructions with almost identical Fermi surfaces, as seen by comparison of Figs.~\ref{fig:1}(g)-(h) and Figs.~\ref{fig:1}(j)-(k), despite their very different magnetic structures in real space.

\section{Nonmagnetic impurity}

In this section we briefly analyze the effects of a non-magnetic impurity introduced by the term
\begin{equation}
 \mathcal{H}_p=V_p \sum_{\mu\sigma}c_{\mathbf{i^*} \mu\sigma}^{\dagger}c_{\mathbf{i^*} \mu\sigma},
\end{equation}
which adds a local spin-less potential at site $\mathbf{i^*}$. 
The orbitally diagonal potential $V_p$ is a good approximation in these systems~\cite{Snakamura}.  
We calculate the projected spin resolved LDOS
\begin{align}
 N_{\sigma\sigma'}(\mathbf{i})&=-\frac{1}{\pi} \mbox{Im}(\mathcal{G}_{\sigma\sigma'}(\mathbf{i},\omega))=-\frac{1}{\pi} \mbox{Im}\sum_{n,\mu}\frac{u_{\mathbf{i}\mu\sigma}^n u_{\mathbf{i}\mu\sigma'}^n}{\omega-E_n+i\eta}, \\\nonumber
\end{align}
to get the total LDOS $N(\mathbf{i})=\sum_{\sigma\sigma'}N_{\sigma\sigma'}(\mathbf{i})$, and the local spin-polarization of the electrons at the Fermi energy ($\omega=0$), $\mathbf P(\mathbf{i})=\textrm{Tr}\left( \mathbold\sigma_{\sigma\sigma'} N_{\sigma\sigma'}(\mathbf{i})\right)/N(\mathbf{i}))$.

\begin{figure}[t]
\begin{center}
\includegraphics[width=0.6\columnwidth]{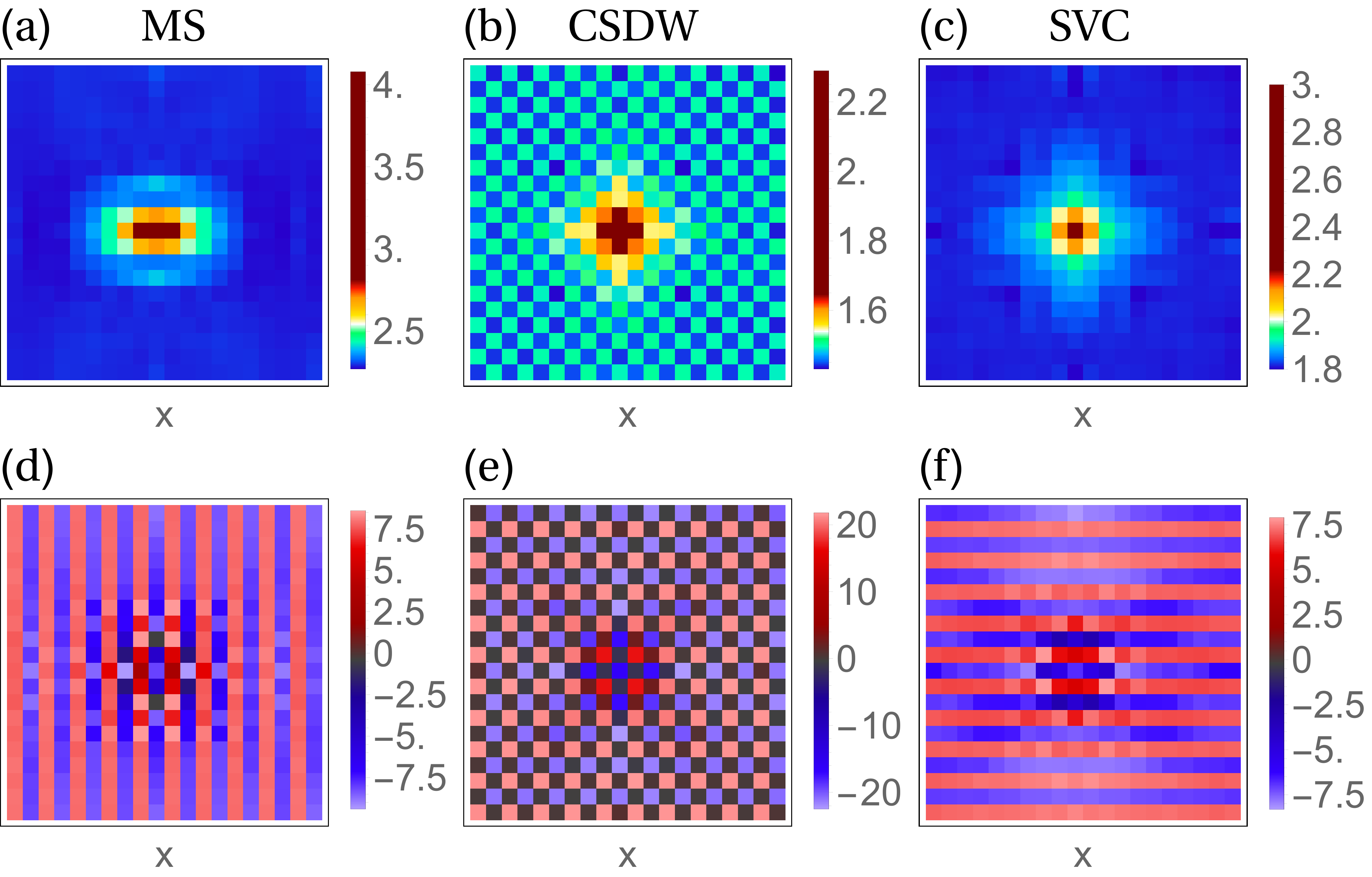}
\end{center}
\caption{(a.c) Total (spin-summed) LDOS $N(\mathbf i)$ at $\omega=0$ around a $V_p=0.5$ eV non-magnetic impurity in the (a) MS, (b) CSDW, and (c) SVC states.
(d)-(f) Local spin polarization percent $\mathbf P(\mathbf{i})\cdot \mathbf{\hat l} $ $[\%]$ of the corresponding cases in the upper row with $\mathbf{\hat l}=\mathbf{\hat z},\mathbf{\hat z}$ and $\mathbf{\hat x}$, respectively.}
\label{fig:2}
\end{figure}

Figures~\ref{fig:2}(a)-(c) display the resulting LDOS at $\omega=0$ around a $V_p=0.5$ eV potential placed in each of the three different magnetic states.
In the MS state [Fig.~\ref{fig:2}(a)] the impurity reflects the broken $C_4$ symmetry of the homogeneous system. By contrast, the tetragonal symmetry is preserved around the potential in both double-$\mathbf{Q}$ phases as seen from Figs.~\ref{fig:2}(b) and \ref{fig:2}(c).
In the CSDW case there is a $(\pi,\pi)$ modulation in $N(\mathbf{i})$, arising from the charge order at $\mathbf q=\mathbf{Q_1}+\mathbf{Q_2}$ of the homogeneous state.
The $(\pi,\pi)$ charge modulation constitutes a strong STM fingerprint of the CSDW ordered state.
The amplitude of this modulation, however, may be too small to be easily detected by tunneling spectroscopy.
This implies that there is no simple way to distinguish the SVC and CSDW phases by use of a non-magnetic potential scatterer.

The local spin polarization $\mathbf P(\mathbf{i})$ gives complementary information that would, in principle, allow one to distinguish between the CSDW and the SVC phases from the spin-polarized tunneling conductance. 
We show in Figs.~\ref{fig:2}(d)-(f) the polarization at the Fermi level for the relevant spin projection $\mathbf{\hat l}$ of the three magnetic states ($\mathbf P(\mathbf i)\cdot\mathbf{\hat l} \propto \mathbf M_{\mathbf i}\cdot\mathbf{\hat l}$).
This property is related to the magnetic contrast measured in a spin-polarized STM experiment \cite{Swiesendanger09}.
The polarization of the single-$\mathbf{Q}$ state in Fig.~\ref{fig:2}(d) consists of $\mathbf{Q_1}$ modulated stripes for the $\mathbf{\hat l}=\mathbf{\hat z}$ projection.
The $\mathbf{\hat l} =\mathbf{\hat x} $ and $\mathbf{\hat l} =\mathbf{\hat y} $ components have no polarization, since $\mathbf M_{\mathbf i}\cdot \mathbf{\hat l} =0$ in the $xy$ plane [Fig.~\ref{fig:1}(c)].
In the case of the CSDW state, the relevant projection is also the $\mathbf{\hat z}$ axis [Fig.~\ref{fig:1}(d)], where half of the sites appear with alternating polarization and the other half are not polarized, resulting from an equal superposition of $\mathbf{Q_1}$ and $\mathbf{Q_2}$ parallel spin density waves. Note that the $C_4$ symmetry is preserved in the polarization pattern around the impurity in this state.
The last magnetic state, the SVC, with a coplanar spin structure [Fig.~\ref{fig:1}(e)] has two relevant spin projections, the in-plane $\mathbf{\hat l}=\mathbf{\hat x}$ and $\mathbf{\hat l}=\mathbf{\hat y}$.
$\mathbf{Q_2}$ modulated stripes can be seen in Fig.~\ref{fig:2}(f) for $\mathbf{\hat l}=\mathbf{\hat x}$, with a local $C_2$ symmetric polarization pattern around the impurity.
The perpendicular $\mathbf{\hat l}=\mathbf{\hat y}$ polarization (not shown) consists of $\mathbf{Q_1}$ stripes, similar to those in the MS state [Fig.~\ref{fig:2}(d)].
The two double-$\mathbf{Q}$ phases are now clearly discernible, with distinctive local and global spin-polarization tunneling.


\end{document}